\documentclass[aps,prl,preprint]{revtex4}
\usepackage{graphicx}
\usepackage{verbatim}

\newcommand{\req}[1]{(\ref{#1})}
\newcommand{\be}{\begin{equation}}
\newcommand{\ee}{\end{equation}}
\newcommand{\bea}{\begin{eqnarray}}
\newcommand{\eea}{\end{eqnarray}}

\newcommand{\cro}[1]{\left[#1\right]}

\newcommand{\avg}[1]{\langle{#1}\rangle}

\newcommand{\BE}{\begin{eqnarray}}
\newcommand{\EE}{\end{eqnarray}}
\newcommand{\BEn}{\begin{eqnarray*}}
\newcommand{\EEn}{\end{eqnarray*}}
\newcommand{\barr}{\begin{array}}
\newcommand{\earr}{\end{array}}

\newcommand{\bit}{\begin{itemize}}      
\newcommand{\eit}{\end{itemize}}
\newcommand{\bc}{\begin{center}}
\newcommand{\ec}{\end{center}}
\newcommand{\ben}{\begin{enumerate}}    
\newcommand{\een}{\end{enumerate}}

\newcommand{\eps}{\epsilon}

\begin{document}

\title{The demise of constant price impact functions and single-time step models of speculation}
\author{Damien Challet}
\affiliation{Nomura Centre for Quantitative Finance, Mathematical Institute, Oxford University, 24--29 St Giles', Oxford OX1 3LB, United Kingdom}
\email{challet@maths.ox.ac.uk}

\date{\today}


\begin{abstract}
Constant and symmetric price impact functions,  most commonly used in agent-based market modelling, are shown to give rise to paradoxical and inconsistent outcomes in the simplest case of arbitrage exploitation when open-hold-close actions are considered. The solution of the paradox lies in the non-constant nature of real-life price impact functions. A simple model that includes explicit position opening, holding, and closing is briefly introduced and its information ecology discussed, shedding new light on the relevance of the Minority Game to the study of financial markets.
\end{abstract}

\maketitle

The immense majority of agent-based financial market models relies on two fundamental assumptions: constant and often symmetric price impact function, and agents with a strategy time-horizon of a single time step (see e.g. \cite{CB,Lux,BG01,dollargame,Alfarano,Hommes_HAM_review,CM03}). Both assumptions are useful simplifications that made possible some understanding of the dynamics of such models. But the modelling of financial markets crucially depends (or should depend) on the fact that earning or losing money occurs while doing nothing, that is, while holding a position. This is to be compared with mainstream agent-based models where the agents' gain is the result of transactions. Since these models are generally defined in discrete time, one could in principle argue that one time step is long enough to include holding periods, but this is inconsistent with the nature of most financial markets. Indeed, the buy/sell orders arrive usually asynchronously in continuous time, which rules out the possibility of synchronous trading; even more, the order in which the limit orders are submitted raises important questions about the nature of financial markets \cite{C05,SolomonMarkovNet}.

\section*{The problem with constant price impact functions}
Let me start with the price impact function. Such a function $I(n)$ is by definition the relative price change induced by a transaction of $n$ (integer) shares; mathematically, 
\be
p(t+1)=p(t) + I(n),
\ee
where $p(t)$ is the {\em log}-price and $t$ is the transaction time. The above notation misleadingly suggests that $I$ does not depend on time. In reality, not only $I$ suffers from random fluctuations, but also, for instance, from strong feed-back from the sign of the last transactions, which has long memory (see e.g. \cite{BouchaudLimit3,FarmerSign,BouchaudLimit4} for discussions about the dynamical nature of market impact). Neglecting the dynamics of $I$ requires to consider specific shapes for $I$ that enforce some properties of price impact for each transaction, whereas they should only hold on average. For example, one should restrict oneself to the class of functions that make it impossible to obtain round-trip positive gains \cite{FarmerForce}. But the inappropriateness of constant price impact functions is all the more obvious as soon as one considers how price predictability is removed by speculation, which is inter-temporal by nature.

The most intuitive (but wrong) view of market inefficiency is to regard price predictability as a scalar deviation from the unpredictable case: if there were a relative price deviation $r_0$ caused by a transaction of $n_0$ shares at some time $t$, according to this view, one should exchange $n_1$ shares so as to cancel perfectly this anomaly, where $n_1$ is such that  $I(n_1)=-r_0$. This view amounts to regard predictability as something that can be remedied with a single trade. However, the people that would try and cancel $r_0$ would not gain anything by doing it, unless they are market makers who try to stabilise the price. The speculators on the other hand make money by opening, holding, and closing positions. Hence one needs to understand the mechanisms of inefficiency removal by the speculators.

The following discussion is best understood if one assumes that there are at first no transactions but that of trader $0$.\footnote{One can equivalently assume that the speculators experience no difficulties in injecting their transactions just before and after that of trader $0$.} A perfectly (and probably illegally) informed speculator (called trader $1$ thereafter) will take advantage of his knowledge by opening a position at time $t-1$ and closing it at time $t+1$. It is important to be aware that if one places an order at time $t$, the transaction takes place at $p(t+1)$. The round-trip of trader $1$ yields a monetary gain of
\be
g_1=n_1[e^{p(t+2)}-e^{p(t)}]=n_1e^{p_0}[e^{I(n_0)}-e^{I(n_1)|}].
\ee
where $p_0$ is the price before any trader considered here is active.
Since $I(n)$ generally increases with $n$, there is an optimal $n_1^*$ number of shares  that maximises $g_1$. For obvious reasons, it is advantageous to consider $I(x)=\log f(x)$ where $f$ is an odd function of $x$; the average real-life shape of $I$ is a concave function described by $I(x)=x^\alpha$ or $I(x)=\log x$ \cite{FarmerImpact,BouchaudLimit2}. The following discussion focusses on $f(x)=\textrm{sign x}|\lambda x|^\gamma$; the presence of $\lambda>1$ ensures that a transaction of one share results in a price change; for a while only, $n_2$ is rescaled so as to contain $\lambda$, which shorten the mathematical expressions. The optimal number of shares to invest is therefore
\be
n_1^*=\frac{n_0}{(\gamma+1)^{1/\gamma}},
\ee
which simplifies to $n_0/2$ if $\gamma=1$. The optimal gain is given by
\be\label{eq:g1opt}
g_1^*=e^{p_0}n_0^{\gamma+1}\frac{\gamma }{(\gamma+1)^{1+1/\gamma}}
\ee
The discussion so far is a simplification, in real-money space,  of the one found in Ref. \cite{FarmerForce}. One should note that far from diminishing price predictability, the intervention of trader $1$ increases the fluctuations. Therefore, in the framework of constant price impact functions, predictability never vanishes but becomes less and less exploitable because of the fluctuations and the reduction of signal-to-noise ratio caused by the speculators. The other obvious consequence of the activity of trader $1$ is the additional cost for trader $0$ since the price he obtains is increased by a factor $e^{I(n_1)}$.

It seems that trader $1$ cannot achieve a better gain than by holding $n_1^*$ shares at time $t$. However, he can inform a fully trustable friend, trader $2$, of the gain opportunity on the condition that the latter opens his position before $t-1$ and closes it after $t+1$ so as to avoid modifying the gain of trader $1$. For instance, trader $2$ informs trader $1$ when he has opened his position and trader $1$ tells trader $2$ when he has closed his position. From the point of view of trader $2$, this is very reasonable because the resulting action of trader $1$ is to leave the arbitrage opportunity unchanged to $r_0$ since $p(t+1)-p(t-1)=r_0$. Trader $2$ will consequently buy $n_2^*=n_1^*$ shares at time $t-2$ and sell them at time $t+2$, earning the same amount of money as trader $1$. This can go on until trader $i$ has no fully trustable friend. Note that the advantage of trader $1$ is that he holds a position over a smaller time interval, thereby increasing his return rate. Before explaining the paradox of this situation, it makes sense to emphasise that the gains of trader $i>0$ are of course obtained at the expense of trader $0$, and that the result the particular order of the traders' actions is to create a bubble which peaks at time $t$.

The paradox is the following: if trader $1$ is alone, the most that can be extracted from his perfect knowledge is $g_1(n_1^*)$ according to the above reasoning. When there are $N$ traders in the ring of trust, the total money extracted is $N$ times the optimal gain of a single trader. Now, assume that trader $1$ has two brokering accounts; he could play with each of his accounts, respecting the order in which to open and close his positions, effectively earning $g_1(n_1^*)$ on each of his accounts. The paradox is that his actions would be completely equivalent to investing $n_1^*$ and then $n_1^*$ from the same account. In the case of $I(n)=n$, this seems {\em a priori} exactly similar to grouping the two transactions into $2n_1^*$, but this results of course in a gain smaller than $g_1(n_1^*)$ for a doubled investment. Hence, in this framework, trader $1$ can earn as much as it pleases him provided that he splits his investment into sub-parts of $n_1^*$ shares.

Two criticisms can be raised. First, the ring of trust must be perfect for this scheme to work, otherwise a Prisoner's dilemma arises, as it is advantageous for trader $i+1$ to defect and close his position before trader $i$. In that case, the expected payoff for each trader is of order $1/N$, as in Ref \cite{FarmerForce}.

But more importantly, one may expect that the above discussion depends crucially on the fact that $r_0$ does not depend on the actual price, or equivalently that trader $0$ wishes to buy or to sell a pre-determined number of shares. One must therefore examine the situation where trader $0$ has a fixed budget $C$ (which modifies the discussion only if trader $0$ intends to buy); he can buy $n_0^{(1)}=C/e^{p_0+I(n_1)}=n_0/e^{I(n_1)}$ shares of the asset, and the price return is now $r_0^{(1)}=r_0/e^{I(n_1)}$. 
The optimal number of shares for player $1$ is 
\be
n_1^*=\cro{n_0^\gamma(1-\gamma)}^{\frac{1}{\gamma(\gamma+1)}};
\ee
in particular, $n_1^*=0$ if $\gamma=1$: the maximum gain is achieved by trading as little as possible; this is due to the peculiar choice of $f(x)$, which cancels exactly the term of order $n_1$ in $g_1$. Since the real-life shape $I$ is not known with certainty, it makes sense to consider other choices of $\gamma$. As soon as $\gamma<1$, $n_1^*>0$. The optimal gain is given by
\be
g_1^*=e^{p_0}n_0\frac{\gamma}{(1-\gamma)^{1-1/\gamma}},
\ee
which is linear in $n_0$, in constrast with its super-linearity in Eq \req{eq:g1opt}: the limited budget of trader $0$ helps to remove predictability.

It is not anymore the interest of trader $1$ to communicate for free the existence of the arbitrage opportunity to one of his friends,  because the action of trader $2$ would reduce the price bias $r_0$ of trader $0$ to $r_0^{(2)}=r_0^{(1)}/e^{I(n_2)}$; he would therefore loose
\be
\Delta g_1(n_2)=g_1^*-g_1(n_1^*,n_2)=e^{p_0}n_0(1-\gamma)^{1/\gamma-1}\cro{1-\frac{1}{n_2^\gamma}}
\ee
\begin{figure}
\centerline{\includegraphics[width=0.5\textwidth]{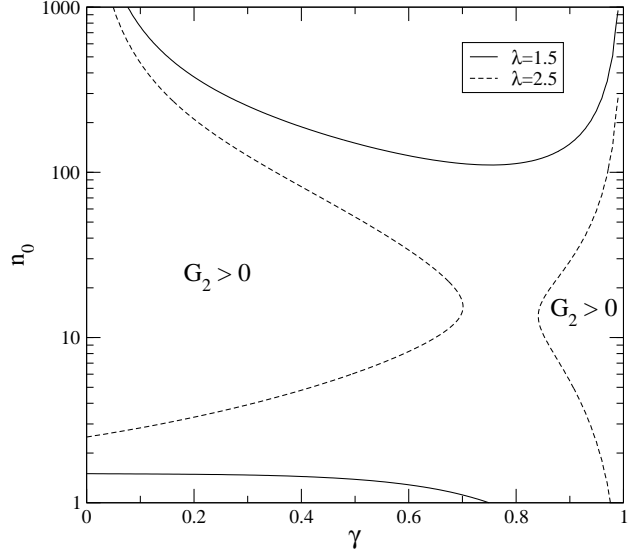}}
\caption{Regions of positive and negative gain of trader $2$ in parameter space ($n_0$, $\gamma$) }
\label{fig:g2tilde}
\end{figure}
His alternative is to sell this information for a sum $\Delta g_1^*$ that compensates for the reduction of arbitrage caused by trader $2$. Therefore trader $2$ optimises
\be
\frac{G_2}{e^{p_0}}=[g_2-\Delta g_1]e^{-p_0}=\cro{\frac{n_0}{1-\gamma}}^{\frac{\gamma}{(\gamma+1)}}n_2^{1-\gamma^2}-n_2^{1+\gamma}-n_0(1-\gamma)^{1/\gamma-1}\cro{1-\frac{1}{n_2^{\gamma^2}}}
\ee
 with respect to $n_2$. The paradox survives in the regions of the parameter space such that $G_2>0$; an early indication that it is possible comes from setting $n_2=1$, which gives $n_0>1-\gamma$: in this case, the paradox is still valid at least for $n_2\le 1$. It is not possible to solve analytically the extremalisation of $G_2$ with respect to $n_2$ but  numerical investigations show that the first derivative of $G_2$ is always negative, thus, it is best for trader $2$ to invest the smallest amount possible. One fixes therefore $n_2$ to 1; at this stage, it is beneficial to reintroduce $\lambda$ and study where $G_2>0$ in the  space ($\gamma$, $n_0$) as a function of $\lambda$. Fig \ref{fig:g2tilde} plots the dividing line between positive and negative gains for trader $2$. As expected, $G_2>0$  requires a large enough $n_0$; more surprising is the fact that $G_2<0$ if $n_0$ is too large, which is entirely due to the cost of information $\Delta g_1$. When $\lambda\ge\lambda_{c}\simeq 2.4495$, the upper and lower regions merge, and positive $G_2$ are only found on the left and right hand side of the plane. Both regions gradually shrink when $\lambda$ increases further, the right hand side one becoming very small for $\lambda>10$. The left hand side region of positive gain gradually shrinks and shifts towards higher values of $n_0$, but is  still of a fair size for $\lambda=50000$.

The solution to this paradox lies in the feed-back from past actions onto the order book. For instance, although the sign of market orders has a long memory \cite{FarmerSign,BouchaudLimit3}, the market is still efficient because the order book's reaction prevents statistical arbitrage, decreases notably the exploitability of $r_0$.\footnote{A full discussion of this point will be reported elsewhere.}

\section*{A simple strategy space for modelling intertemporal speculation}

An important point raised by the above discussion is the need to consider markets in transaction time if one wishes to understand their minute dynamics, which in turn begs for models where the agents' actions are inter-temporal and intertwined. The design of simple models that allow the agent to open, hold, and close their position explicitely faces two problems. First, as explained above, a speculator makes money when holding a position; therefore a trader that has an open position wishes to know for how long he should hold it, or equivalently, when to close it. Given the amount of randomness in the markets, one is only interested in a criterion based on statistical analysis of market history; in other words, this criterion should only be trusted on average. The next problem to find a trading strategy space that has as few parameters as possible. One possibility is to study strategies reportedly used by practitioners as in Ref. \cite{FarmerJoshi}; they usually prescribe very concretely when to open and close a position as a function of the current price and of other parameters. Another possibility is to take inspiration from the Minority Game and to  look for a strategy space based on states of the world (which might include statistics about past prices) that contains a finite and countable number of elements; in that way, one still expects the behaviour of the system to change significantly when the number of agents exceeds the effective number of available strategies.

It is reasonable to assume that real-life traders base their decisions on signals or patterns, such as mispricing (over/undervaluation), technical analysis, crossing of moving averages, news, etc. How to close a position is a matter of more variations: one can assume a fixed-time horizon, stop gains, stop losses, etc. I assume that the traders are inactive unless they receive some signal because some known pattern arises. Therefore, the agents hold their positions between two recognisable patterns. All the kinds of information regarded as possibly relevant by all the traders form the ensemble of the patterns which is assumed to be countable and finite. Mathematically, the state of the world is fully characterised by an integer number $\mu\in\{1,\cdots,P\}$.

Each trader recognises only a few patterns, because he has access to or can analyse only a limited number of information sources, or because he does not regard the other ones as relevant; in the present model, a trader is assigned at random a small set of personal patterns which is kept fixed during the whole duration of the simulation. Every time one of his patterns arises, he decides whether to open/close a position according to his measure of the average return between all the pairs of patterns that he is able to recognise. This is precisely how people using crossings of moving averages behave: take the case of a trader comparing two exponential moving averages (EMA) EMA100 and EMA200, with memory of 100, respectively 200 days: such trader is inactive unless EMA100 and EMA200 cross; the two types of crossing define two signals, or patterns. For instance, a set of patterns could be the 20 possible crossings between EMAs with memory length of 10, 20, 50, 100 and 200 days.

 The hope and the role of a trader are to identify pairs of patterns such that the average price return between two patterns is larger than some benchmark, for instance a risk-free rate (neglecting risk for the sake of simplicity); in this sense the agents follow the past average trend {\em between two patterns}, which makes sense if the average return is significantly different from the risk-free rate. In contrast with many other models (e.g. Ref \cite{Lux}), the agents do not switch between trend-followers/fundamentalist/noise traders during the course of the simulation.

\begin{figure}
\centerline{\includegraphics*[width=0.7\textwidth]{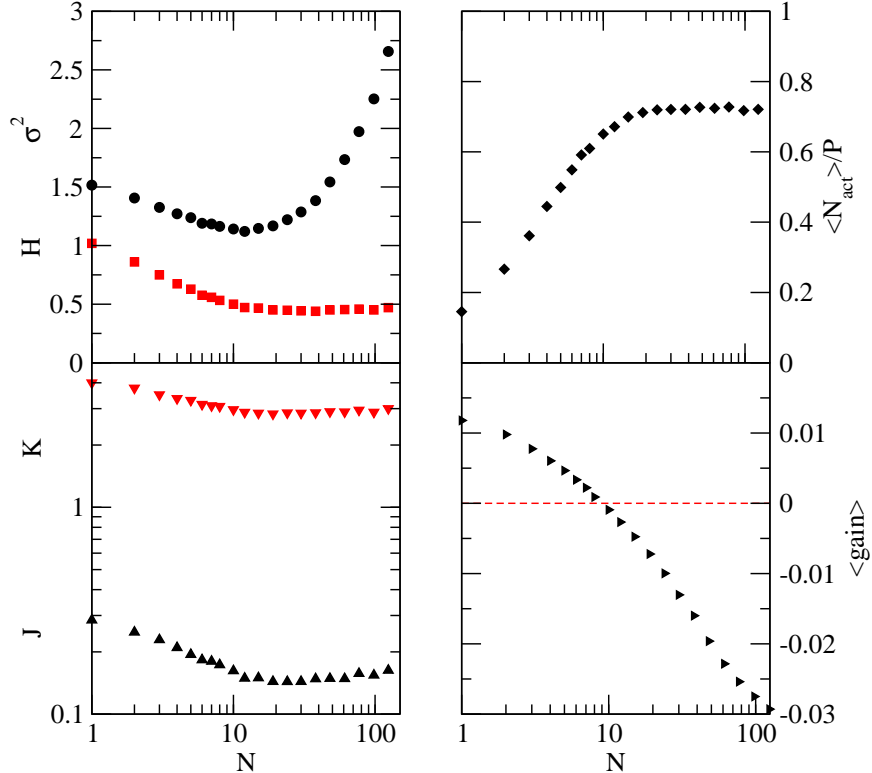}}
\caption{Price volatility $\sigma^2$ (circles), price impact $H/P$ (squares), naive predictability $J$ (down triangles) and sophisticated predictability $K$ (up triangles), scaled fraction of active speculators (diamonds), and average gain per speculator and per time step (right triangles); $P=5$, $N_p=10$, $\eps=0.05$; average over 10000 samples of speculators}
\label{fig:ecology-4figs}
\end{figure}

The full definition of the model can be found elsewhere \cite{C05}.\footnote{Somehow ironically, this model uses constant price impact functions} A most valuable consequence of the way the model is defined is that, similarly to the Minority Game, one can introduce a well-controlled amount of predictabily and study how some agents, named speculators remove it. The similarity of the interplay between price fluctuations and predictability in this model and in the Minority Game is striking (Fig. \ref{fig:ecology-4figs}): the more speculators one adds, the less predictable the market ($J$ and $K$, lower left panel), and also the smaller the price fluctuations ($\sigma^2$, upper left panel); in addition, the average number of active speculators saturates, and their gain, positive when they are only a few, decreases and becomes negative when they are too many. The similarity stops there, since the predictability seems impossible to completely remove with the kind of strategies available to the agents; more precisely, the agents do not seem to be able to manage to remove more than a fraction of it. Nevertheless, the above similarity substantiates the use of the MG as a model of information ecology, and suggests the need to re-interpret it.

Generally speaking, a minority mechanism is found when agents compete for a limited resource, or equivalently when they try to determine collectively by trial and error the (sometimes implicit) resource level, and synchronise their actions so that the demand equals the resource on average \cite{CMO03,C04}. As a consequence, an agent is tempted to exploit the resource more than his fair share, hoping that the other agents happen to take a smaller share of it. The limited resource in financial markets is {\em predictability} and indeed information ecology is one of the most fascinating and plausible insights of minority games into market dynamics \cite{MMM,CM03}. Instead of regarding the difference between the number of agents choosing $+1$ and the ones choosing $-1$, denoted $A(t)$, in the MG as the instantaneous excess demand, one should reinterpret it in a more abstract way as the {\em deviation from unpredictability} $A=0$ at time $t$. The two actions can be for instance to exploit an inefficiency ($+1$) or to refrain from it ($-1$); $A$ in this context would measure how efficiently an inefficiency is being exploited. Then everything becomes clearer: the fact that  $A(t)$ display mean-reverting behaviour is not problematic any more as it is not a price return. It simply means when the agents tend to under-exploit some predictability, they are likely to exploit it more in the next time steps, and reversely. What price return correspond to a given $A$ is not specified by the MG, but herding in the information space (the MG) should translate into interesting dynamics of relevant financial quantities such as volume and price; for instance, dynamical correlations of $|A|$ in the MG probably correspond to dynamical correlations in the volume of transactions. Therefore, studying the building up of volatility feedback, long-range memory and fat-tailed $A$ still makes sense, but not in a view to model explicitly price returns. 

A more formal relationship is that in the MG the state of the world (or history)  $\mu$ in akin to a pair of pattern in the inter-temporal model. What the MG neglects is the interplay between the predictability associated with two pairs of patterns, or put differently, the intrisically inter-twined nature of transactions and predictability. Mathematically  the dependence of the average price returns between $\mu_1$ and $\mu_2$ and, for instance, between $\mu_1$ and $\mu_3$ is simply replaced by two independent quantities $\avg{A|\mu'}$ and $\avg{A|\mu''}$, in the standard notation of the MG.

\section{Conclusions}

In conclusion, dropping the assumption of constant price impact functions and exploring the world of inter-temporal speculation will bring new and fascinating insights on financial market dynamics. One should aim to find an inter-temporal model amenable to mathematical analysis, or a mathematical analysis of the model discussed above.

\bibliography{biblio}










\end{document}